\documentclass[superscriptaddress,aps,amsfonts]{revtex4}
\usepackage{epsfig}
\usepackage{graphicx}
\usepackage{subfigure}
\usepackage{color}

\begin{document}
\title{XPM nonlinearities:Are all involved photons real? }
%\shorttitle{ Multiphoton processes due to XPM nonlinearities in EIT systems.} 
\author{Archana  Sharma}
%\inst{1} \thanks{E-mail:\email{pegionhole@gmail.com}}}
\affiliation{LAMP dept.,Raman Research Institute,  Sadashiva Nagar, Bangalore - 560
080, India,E-mail:pegionhole@gmail.com}

%\pacs{42.50.Gy}{Effects of atomic coherence on propagation; absorption,and amplification of light; electromagnetically induced transparency and absorption}

%\pacs{42.50.Hz}{strong field excitation of optical transitions in quatum systems; multiphoton processes; dynamic stark shift}

%\pacs{42.50.Lc}{Quantum fluctuations,quantum noise,and quantum jumps}

\begin{abstract}
 Many symmetries of nonlinear susceptibilities\cite{boyd},especially in crystals,are well known.We have found a novel symmetry for XPM(Cross Phase Modulation) systems .We prove that this symmetry is a necessary  condition for a single multiphoton process to be visible in more than one field.We have found, two EIT(Electromagnetically Induced Transparency) systems that show cases of this symmetry.The above process has the possibility of leading to strong sources of temporally entangled photons of different frequency.
For cases that do not show this symmetry,more than one "species" of multiphoton processes can take place and each one is visible in one field only.
%However,an asymmetric case might be seen as a sum of the symmetric and   
% asymmetric part.
%For this reason we agree only partially with Harris\cite{harris} that the equality, which is stated by him without giving any reasons, is a symmetry.
 % Whether or not the symmetry is a sufficient condition for a single   
 %multiphoton process to be visible in more than one field,
When there is a symmetry, we  predict cross field noise correlations for the scattered fields.Cross field noise correlations have recently been seen in a different situation that can provide interesting variation to our study,namely the generated fields in FWM(Four Wave Mixing) \cite{xiao},\cite{maria},\cite{boyer}.
We postulate or conjecture the existence of two types of asymmetric multiphoton processes. We show only one of these is possible.
One of systems we use for this purpose generates a $\chi^{(9)}$ absorption resonance that is observable on the same trace that shows  linear features, in an experiment described in a previous work\cite{andal} but the order of nonlinearity newly interpreted using this work.This high order nonlinearity should be very sensitive to cross field correlations.This can be used to test our conjecture of asymmetric processes. 
 %From the two systems we analyze,we show that there are two ways in which a 
%multiphoton process can be rendered invisible in all fields save one.Hence, %we see three types of multiphoton processes.
 We also come across an important $\chi^{(5)}$ absorption,comparable to linear absorption, that is anomalous.The anomaly is that the Imaginary part of this XPM nonlinearity can cause squeezing.We give a qualitative argument which shows that we don't expect the Imaginary part of the XPM nonlinearities we usually come across, to cause squeezing.This promising source of squeezing, is qualitatively different from the other sources hitherto known.

\textbf{PACS 42.50.Gy,42.50.Hz,42.50.Lc}
%{Quantum fluctuations,quantum noise,and quantum jumps}

\end{abstract}

\maketitle
 
 \section{Introduction\label{sec:intro}}
\begin{figure}
\includegraphics[scale=0.19]{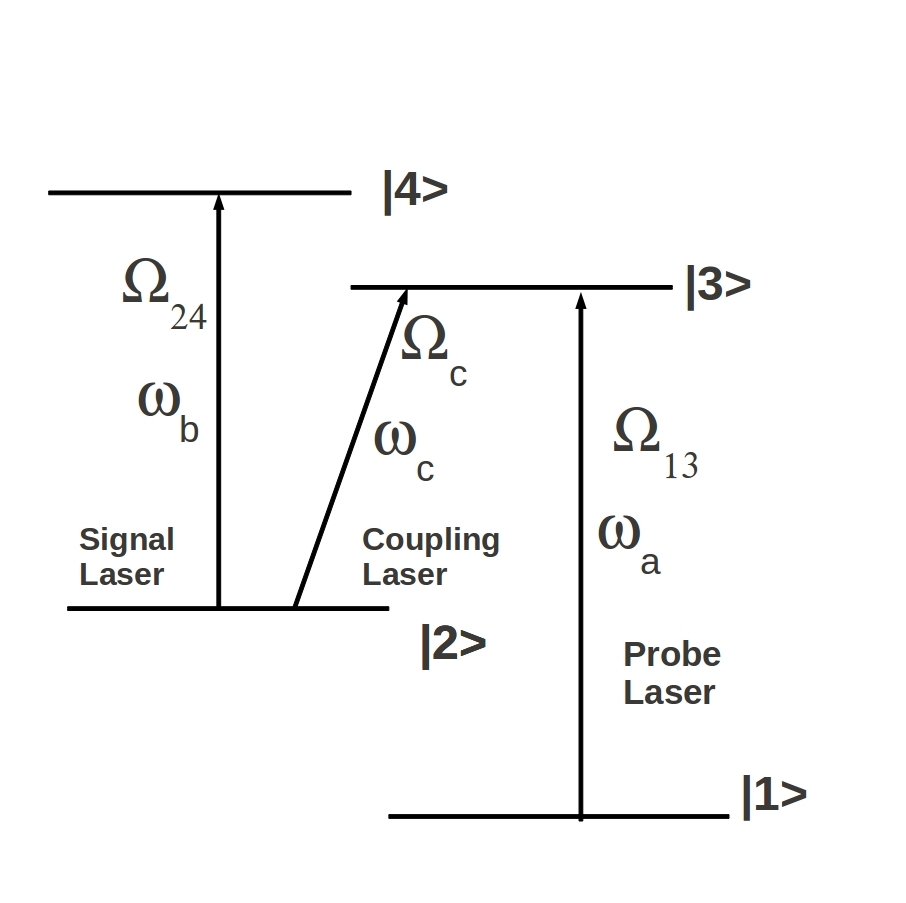} 
\caption{This level scheme represents system1 and has been considered theoretically by Schmidt et.al.\cite{schmidt}.This can represent the D2 line of
$Rb^{87}$ system given, for example, in \cite{kang}.}
%The level scheme on the right is system2 and has been considered by Harris et. al.}
\label{fig:atom-lasersys1}
\end{figure}

\begin{figure}
\includegraphics[scale=0.19]{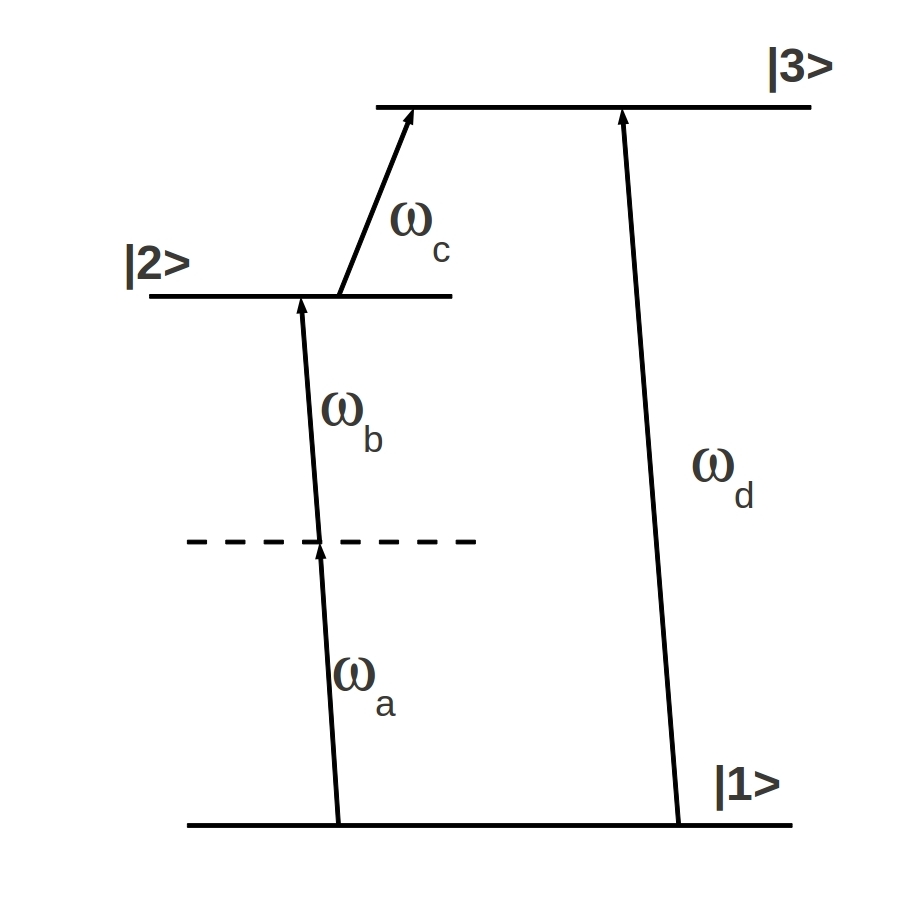} 
\caption{
%The level scheme on the left represents system1.This can represent the D2 line of
%$Rb^{87}$ system as given in \cite{kang}.
%Transitions between levels $|1\rangle$ and $|2\rangle$;$|3\rangle$ and $|4\rangle$ are
%dipole  forbidden.Transitions from $|1\rangle$ to $|4\rangle$ are also dipole
%forbidden.The two beams, viz.,the Signal and the Coupling cannot provide the
% probe coupling because the ground levels $|2\rangle$ and $|1\rangle$ are
 % separated by 
 %a large energy compared to the excited level manifold.
This level scheme is system2 and has been considered theoretically by Harris et. al.\cite{harris}}
\label{fig:atom-lasersys2}
\end{figure}

Symmetries are sought after as they make the analysis of a system easier.There is more than one symmetry of nonlinear susceptibilities seen, especially in crystals, like the Kleinman's symmetry.The symmetry we have found is a necessary condition for a single multiphoton process to be visible in more than one field in XPM\cite{gpagrawal} systems.
  The advent of lasers in 1960's produced a lot of research in the field of multiphoton processes, that absorb multiple photons from a single beam, or SPM nonlinearities.Topics like stepwise nonlinear processes\cite{singh} as opposed to the simultaneous process that this work is about,field correlation
sensitivity \cite{gsagarwal},squeezing\cite{paulina} have been studied.
      XPM multiphoton processes became easily accessible after 1990 with the discovery\cite{harris} of enhancement in the presence of EIT, of these nonlinearities.The
advantage of this enhancement is great as it can, in some cases, make them comparable in strength to linear processes at much lower intensities compared to that at which SPM processes are observable. XPM multiphoton processess are
qualitatively different because of effects that arise from the absorption(emission) of
 photons forming the chain, belonging to different fields.This changes properties like squeezing and field correlation sensitivity.In fact, the difference raises a basic question,relevent to many recent studies\cite{xiao},\cite{maria},\cite{boyer},  whether the nonlinear susceptibilities faced by different fields, due to a single "species" of multiphoton process, bear any relation to each other as the quanta of light absorbed by them is related? We address it in this work. We show in section\ref{sec:equalsus} the \textit{generation of cross field correlation} in the different fields as the result of the symmetry that the XPM nonlinear susceptibilities, due to a single "species" of multiphoton process, faced by different fields are the same. 
We have found, from our calculations, cases which show this symmetry 
in the two atom-laser systems shown in the figures [\ref{fig:atom-lasersys1}], [\ref{fig:atom-lasersys2}].

  %The symmetry is seen in the XPM\cite{gpagrawal}
  %nonlinearities that
 %are enhanced in the presence of Electromagnetically Induced Transparency
 %(EIT) and give rise to multiphoton processes. In general, the linear susceptibility also shows an increase along with
% the nonlinear susceptibility at resonance. But at EIT the linear susceptibility
 %for EIT-forming fields is suppressed while the XPM nonlinearities are enhanced.
 %For EIT process the absorption amplitudes to the split excited
 %state destructively interfere. For a multiphoton process, the two pathways
 %formed by the splitting constructively interfere, hence enhancing the XPM
 %nonlinearity in the presence of EIT. We find that while in EIT condition the nonlinearity in the probe beam is enhanced,in CPT condition the nonlinearity in signal beam,and the Real part of nonlinearities in the probe and coupling beams is enhanced. 
The XPM nonlinearity in the probe beam in the system1 described
 in figure [\ref{fig:atom-lasersys1}]  has been observed
 by Kang and Zhu \cite{kang}.
Three out of roughly a dozen of the XPM nonlinearities considered in this work, in systems given by figures [\ref{fig:atom-lasersys1}],  [\ref{fig:atom-lasersys2}] have been long known and have been considered before
theoretically by Schmidt and Imamoglu \cite{schmidt} and Harris et. al. \cite{harris} 
 respectively.Harris et.al.,without giving any reasons, have brought to notice first one of the symmetries. 
%though,from the analysis we have made, we 
% agree only partially about it's actually being a symmetry.In fact,this    example of "partial symmetry" turns out to be useful as it shows how an    asymmetric case can be seen as the sum of symmetric and asymmetric parts.
%Hence we see that EIT enhanced XPM nonlinearities are well known. An important feature of such XPM nonlinearities is that they have large values at low light intensities.
 
   The hitherto well known EIT enhanced XPM nonlinearities are of the type where the polarization is proportional to $\chi^{(x+y+z)} E_1^{x}E_2^{y}E_3^{z}$ where x=1 or 0 and y, z can be greater than 1, where $E_1$ is the field
that is being observed. x=0 or 1, implies that these well known nonlinearities cannot cause light squeezing by using the Imaginary part. There is however a system, we have found, for which x=5,y=z=0. It should be possible to see light squeezing in it,using the Imaginary part of the nonlinearity.

To see that the Imaginary part of every XPM nonlinear susceptibility starting from $\chi(2)$ does not cause squeezing,consider a nonlinearity $\chi^{(2)}E^{2}$ where E is the field in which it is observed.The intensity of the field, recorded in a detector will not scale down linearly because of the nonlinear medium.It scales down slightly more where there is a positive fluctuation in intensity above the mean and scales down less where there is a negative fluctuation in the intensity as is clear from the fact that $11^2-10^2$=21 while $10^2-9^2$=19.This should be a necessary condition for the Imaginary part of a susceptibility to cause squeezing.Now consider a nonlinearity $\chi^{(3)} E_1E_2^{2}$ which is being observed in the field $E_1$.Since the shot noise pattern of field $E_2$ is different from that of $E_1$ the points that are scaled down more and are scaled less do not necessarily correspond to a positive and negative fluctuation respectively of $E_1$.Hence it cannot lead to squeezing.

 Whatever light squeezing, in an atomic medium, that is known to be possible till now relies on an origin qualitatively different from that of the x=5 nonlinearity. Light squeezing due to atomic kerr nonlinearity, for example, relies mainly on $\chi^{(3)}$ nonlinearities due to saturation and power broadening effects in resonant media that is put in a cavity\cite{bachor}, \cite{lambrecht}.It is also different from SPM nonlinearities in that it has large values at low light levels. However, a connection between XPM and squeezing is not unknown\cite{boivin}. A change in photon statistics,with the crucial use of a cavity, by using the Real part of XPM kerr nonlinearity has been shown\cite{imamoglu},\cite{rebic}. In our system, it should be possible to see sub-poissonian statistics of light,using the Imaginary part of the nonlinearity. This implies that this source of squeezing is novel.We show in fig.[\ref{fig:atom-laser7}]that the enhancement of this source due to CPT(Coherent Population Trapping)makes it comparable to the linear absorption.
 %The main concern of this work is multiphoton processes\cite{sellin},\cite{pert} due to EIT enhanced XPM nonlinearities. We consider the two different above mentioned atom-laser schemes and address the question:How exactly nature handles the susceptibility $\chi$ in each beam so that equal quanta of absorption or emission are visible due to a single multiphoton process. It should be notedthat this is an issue with XPM nonlinearities but not with purely SPM(Self Phase Modulation) nonlinearities that for example lead to the well known Second Harmonic Generation. In the simplest scheme for SHG,the absorption of the two photons is from the same beam.

  The question we have posed, and our calculations, lead us to postulate or conjecture three types of multiphoton processes:

Type1:When two or more fields face equal nonlinear susceptibilities, we show it could be the same multiphoton process that is visible in two or more fields simultaneously.It is counterintuitive because while the induced polarizations by each of the two fields can be unequal, the number of photons absorbed or emitted are equal when susceptibilities are equal(we show).

Type2:In the case of unequal susceptibilities, the multiphoton process should be visible only in one field.In the rest of the fields, because of equal number of absorption and simultaneous emissions of photons in the rest of the fields the process is not visible.We can show it's suggested existence in system1.

Type3: the multiphoton process is visible only in one field due to the presence of a simultaneous reverse process that renders it invisible in other fields. We can show ,this possibility is false.
%it's suggested existence using system3 shown in fig[\ref{fig:atom-lasersys3}].

\begin{figure}
\includegraphics[scale=0.19]{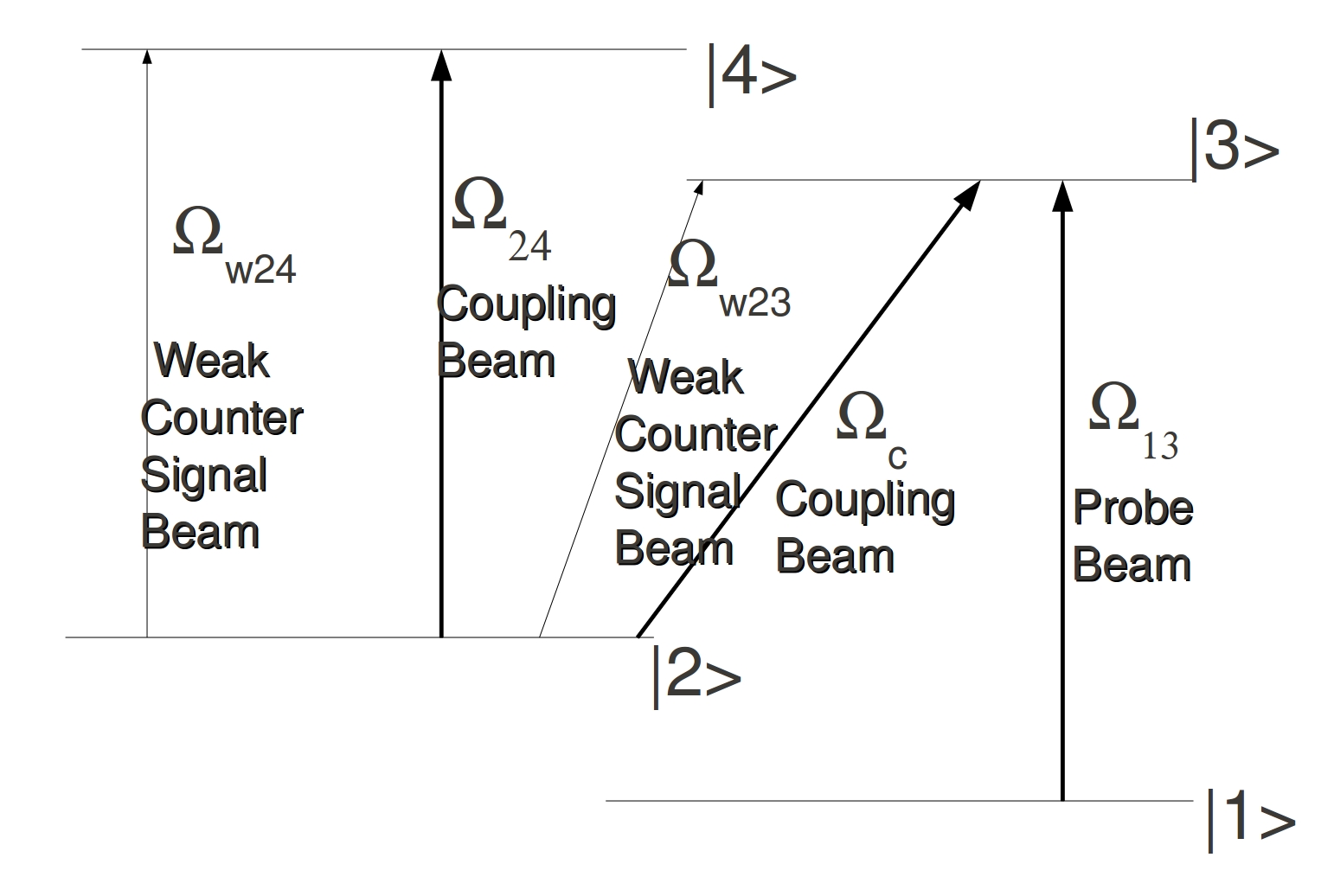} 
\caption{This level scheme represents system3 and is a modified scheme based
on an experiment done before described in \cite{andal}.The weak signal beam
is considered as a perturbation.}
%The level scheme on the right is system2 and has been considered by Harris et. al.}
\label{fig:atom-lasersys3}
\end{figure}

\begin{figure}
\includegraphics[scale=0.59]{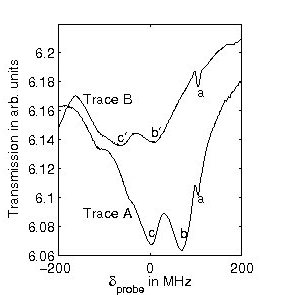} 
\caption{This plot is from an experiment described in our previous work\cite{andal}.The narrow dip shown is newly interpreted as a nine-photon resonance.The other broad dips are linear absorptions.The typical intensity of probe and coupling lasers is about 2mWatt/.04 cmsq. and the weak counter signal beam is 40$\mu$Watt/.04 cmsq..It can be appreciated from this trace that the enhancement of this high order nonlinearity makes it conveniently observable.We expect it to be very sensitive to cross field correlations. }
%The level scheme on the right is system2 and has been considered by Harris et. al.}
\label{fig:exptrace}
\end{figure} 

         The system3 shows a $\chi^{(9)}$ nonlinearity in the beam providing the coupling $\Omega_{w24}$.It is shown in the fig[\ref{fig:exptrace}], obtained from an experiment done earlier\cite{andal}and was described as a three photon resonance  but is newly interpreted as a nine photon resonance using this work.It is shown on the same trace that shows a linear feature.The strength of this high order nonlinearity is important because it is expected to be very sensitive to cross field correlations.The requirement is that there should be photons arriving simultaneously,that belong to different fields.When compared with SPM multiphoton process,this can change the nature of field statistics required for optimum results.XPM processes give a different information about the field statistics.
 A largely asymmetric nonlinearity can be used to experimentally test the conjecture of type2  multiphoton processes through cross field correlation sensitivity. 
%It might tell whether virtual photons are in the picture instead.
\section{The Simulation \label{sec:sim}}
   It is also important to note that the formalism to calculate susceptibilities that has been introduced by Harris et.al. often gives more than one solution. In such cases we either select the valid solution using physical arguments (for system2) or take the solution that matches with density matrix 
simulation\cite{uday}which we have for Schmidt et. al.'s system. 
	     % \section{The Simulation \label{sec:sim}}
  % For the system1 the frequency of
 %the Probe is $\omega_a$,the Coupling is 
 %$\omega_c$ and the
 %Signal is $\omega_b$,.
  %We simulate the four level atom with two lasers acting on it.The Probe and the
  %Coupling.The Coupling
  %laser provides two couplings:one from level $|2\rangle$ to $|3\rangle$ and the
   %other from $|2\rangle$
  %to $|4\rangle$.
The simulation is such that the frequency $\omega_b=\omega_c$. Therefore
  we compare the analytic results, for the special case when the Signal always 
  has same frequency as the Coupling and the Coupling detuning changes, changing
  the position at which the Probe and the Coupling form EIT. We put the changing EIT positions or the coupling beam detuning on the X-axis.Both the analytical results and simulation are for zero velocity of atoms.The coupling detuning is from the upper excited level which is 121 MHz away from the lower excited state.The Probe detuning is from the lower state.
For one case,the simulation is also shown with the ramping probe on the X-axis.

 \section{The XPM nonlinearities in system1 \label{sec:system1}}

 %,from the Schrodinger equation
 %\begin{equation}
 % |\dot{\psi}  \rangle =-\frac{i}{\hbar}H|\psi \rangle
 %\end{equation}

  %where 

%H = \\ \small  {\[ \left |\begin{array}{llcl}  0 &
 % 0 &   -\Omega_{13}\hbar e^{i\omega_a t}/2  &  0   \\   0  &   \hbar\omega_{21}
  %&  -\Omega_{23}\hbar e^{i\omega_c t}/2  &  -\Omega_{24}\hbar e^{i\omega_b t}/2  \\ 
   %-\Omega_{31}^*\hbar e^{i\omega_a t}/2  &  -\Omega_{32}^*\hbar e^{i\omega_c t}/2 

 %& \hbar\omega_{31}-i\Gamma_3/2 & 0 \\
%0 &  -\Omega_{42}^*\hbar e^{i\omega_b t}/2  & 0 & \hbar\omega_{41}-i\Gamma_4/2 
%\end{array}  \right| \]}  
We follow Schmidt et.al.\cite{schmidt} and for
 steady state, after making the rotating wave approximation,
 write the following equations for system1:
 
 \begin{equation}
  -2 \Delta\omega_{21} b_2 +\Omega_c b_3 +\Omega_{24} b_4=0
  \end{equation}

  \begin{equation}
 \Omega_{31}^* b_1 +\Omega_{c}^* b_2 -2A b_3 =0
 \end{equation}

\begin{equation}
 \Omega_{42}^*b_2 -2Bb_4 =0 
 \end{equation}
where  A=$\Delta$$\omega_a$-$\iota$$\Gamma_3$/2
 and B=$\Delta$$\omega_b$-$\iota$$\Gamma_4$/2
 Here $\Omega$'s are the Rabi frequencies and $\omega$'s the frequencies of the
 laser beams and b's are the probability amplitudes of various levels.
 
 \textbf {Case I:when all the atoms are in level $|1\rangle$ or EIT condition}

For this case $b_1^*b_1$=1

 \textit{1.nonlinearity faced by the probe}
 
 \begin{equation}
 b^*_1 b_3 =\frac{-4 \Omega_{13}^* \Delta \omega_{21} A + \Omega_{13}^* 
 |\Omega_{24}|^2}{-2 |\Omega_{24}|^{2} A+8 \Delta \omega_{21} A B-2 |\Omega_c|^{2} B}
 \label{sys1.1}
 \end{equation}
 
 This is the only answer we get for this particular case.
 The polarization(ignoring the oscillations) $Nb^*_1\mu_{13}b_3$ at two photon detuning equal to zero is

 \begin{eqnarray}
&\varepsilon_0& \chi^{(5)} E_{13}E_{24}^4 + \varepsilon_0 \chi^{(5)}
 E_{13}E_{24}^2E_c^2 \nonumber
\\ &=&  \varepsilon_0[\frac{-2N\mu_{13}\mu_{13}^*|\mu_{24}|^4A^*}
 {\varepsilon_0\hbar^5|-2 |\Omega_{24}|^{2} A-2 |\Omega_c|^{2} B|^2}]
 E_{13}E_{24}^4 \nonumber
\\ &+& \varepsilon_0[\frac{-2N\mu_{13}\mu_{13}^*|\mu_{24}|^2|\mu_c|^2B^*}
 {\varepsilon_0\hbar^5|-2 |\Omega_{24}|^{2} A-2 |\Omega_c|^{2}
 B|^2}]E_{13}E_{24}^2E_c^2
\label{probenl} 
%\end{tabbing}
\end{eqnarray}
 
 where $\mu$'s are the dipole matrix elements, and N is the atomic number
density and the terms in square brackets are the fifth order
 susceptibilities, $\chi^{(5)}$. Here,in eq(\ref{probenl}) the second term gives the multi-photon process
$\chi^{(5)}(-\omega_a,\omega_a,-\omega_b,\omega_b,\omega_c,-\omega_c)$, the
emission $-\omega_a$, is out of the beam,thus making the absorption of a $\omega_a$ photon
visible. 
 $\mu_{ab}$ and $\mu_{ab}^*$ mean the absorption and emission of a photon
 respectively or vice-versa.
%Later in this section we tell how to identify whether a particular $\mu_{ab}$ means absorption or emission.     
 
% \begin{figure}
%\includegraphics[scale=0.58]{probeb11.jpg} 
%\includegraphics[scale=0.58]{couplingb11.jpg}
%\includegraphics[scale=0.58]{signalb11.jpg}  
%\caption{The left column of plots is obtained from analytic calculation.The right 
%is obtained from simulation.The parameters are $\Omega_{13}$ = .1  
%  $\Omega_c$=3.55 and $\Omega_{24}$=5.The dashed curve is the dispersion
 % while the solid line is the absorption.The x-axis is the coupling beam detuning 
  %and the y-axis is the probe beam absorption and dispersion in the first row,
  %for Coupling beam in the second row and for Signal beam in the third row.These 
  %plots are for the case when all atoms are in level $|1\rangle$ as can be concluded by
   %looking at the rabi frequencies.}
%\label{fig:caseb11}
%\end{figure}

\textit{2.nonlinearity faced by the Coupling beam can be deduced from the following:}
\begin{equation}
b_2^*b_3=\frac{-\Omega_c^* |\Omega_{13}|^2 |\Omega_{24}|^2B^*}{2BB^*DD^*} + \frac{\Delta
\omega_{21} \Omega_c ^*|\Omega_{13}|^2}{DD^*}
\label{eq:coupling}
\end{equation}
  
  where 
  \begin{equation}
  D=4 \Delta \omega_{21}A - \frac{|\Omega_{24}|^2A}{B} -|\Omega_c|^2
  \end{equation} 
  
\begin{figure}
\includegraphics[scale=0.7]{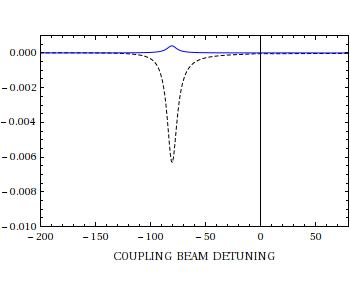} 
\caption{This figure shows the numerically calculated absorption in solid line and dispersion in dashed line, in the coupling beam, in the CaseI.The x-axis is the coupling (or signal beam detuning) beam detuning or the changing EIT position. The parameters are $\Omega_{13}$ = 0.1 MHz,  
  $\Omega_c$=3.55 MHz and $\Omega_{24}$=5 MHz.}
%The level scheme on the right is system2 and has been considered by Harris et. al.}
\label{fig:atom-laser1}
\end{figure}

\begin{figure}
\includegraphics[scale=0.7]{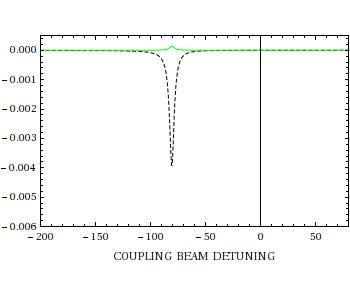} 
\caption{This figure shows the analytically calculated absorption and dispersion in the coupling beam for Case I.The conditions remain the same as in fig[\ref{fig:atom-laser1}].The density matrix simulation treats mixed states while the analytical calculation pure states.The solutions match. }
%The level scheme on the right is system2 and has been considered by Harris et. al.}
\label{fig:atom-laser2}
\end{figure}

This solution matches well with the simulation as shown in fig[\ref{fig:atom-laser1}]and[\ref{fig:atom-laser2}]  
  
  \textit{3.nonlinearity faced by the Signal beam can be deduced from the following:}
\begin{equation}
b_2^*b_4=\frac{|\Omega_c|^2 |\Omega_{13}|^2 \Omega_{24}^* B^*}{2BB^*DD^*}
\label{sys1.3} 
\end{equation}
  
\begin{figure}
\includegraphics[scale=0.7]{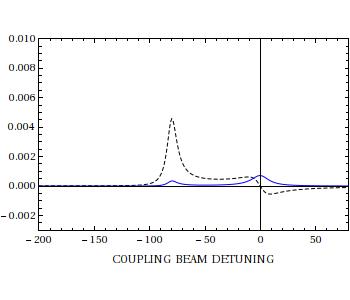} 
\caption{This figure shows the numerically calculated absorption and dispersion in the Signal beam for Case I.The conditions remain the same as in fig[\ref{fig:atom-laser1}].The density matrix solution shows a linear absorption at detuning=0.For the analytic solution given in fig[\ref{fig:atom-laser4}],$b_1b_1^*$=1,so there is no population in level $|2\rangle$,because of which there is no linear part. The rest of the solution matches. }
%The level scheme on the right is system2 and has been considered by Harris et. al.}
\label{fig:atom-laser3}
\end{figure}

\begin{figure}
\includegraphics[scale=0.7]{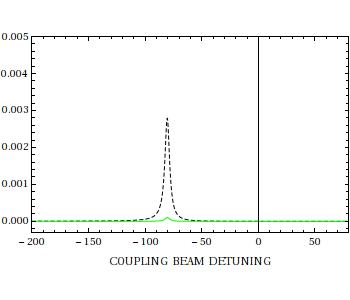} 
\caption{This figure shows the analytically calculated absorption and dispersion in the signal beam for Case I.The conditions remain the same as in fig[\ref{fig:atom-laser1}]. }
%The level scheme on the right is system2 and has been considered by Harris et. al.}
\label{fig:atom-laser4}
\end{figure}

 The above solution matches well with simulation as shown in fig[\ref{fig:atom-laser3}]and[\ref{fig:atom-laser4}]   
 %\textbf {Case II:when all the atoms are in level $|2\rangle$}

 %  For this case $b_2^*b_2$=1 

% \textit{1.nonlinearity faced by the probe}

 % \begin{equation}
  %b_1^*b_3=\frac{2\Delta \omega_{21}D^*}{|\Omega_c|^2\Omega_{13}} - \frac{2|
  %\Omega_{24}|^2 D^*}{4B |\Omega_c|^2 \Omega_{13} }
  %\end{equation} 
 
 %\textit{2.nonlinearity faced by the Coupling beam}
 %\begin{equation}
 %b_2^*b_3=\frac{2\Delta \omega_{21}}{\Omega_c}-\frac{|\Omega_{24}|^2}{4B}
 %\end{equation}
 
 %\textit{3.nonlinearity faced by the Signal beam}
 %\begin{equation}
 %b_2^*b_4=\frac{\Omega_{24}^*}{2B}
 %\end{equation}
 
 %\begin{figure}
%\includegraphics[scale=0.58]{probeb22.jpg} 
%\includegraphics[scale=0.58]{couplingb22.jpg}
%\includegraphics[scale=0.58]{signalb22.jpg}  
%\caption{The left column of plots is obtained from analytic calculation.The right 
%is obtained from simulation.The parameters are $\Omega_{13}$ = 3.55  
 % $\Omega_c$=.1 and $\Omega_{24}$=.1.The dashed curve is the dispersion
  %while the solid line is the absorption.The x-axis is the coupling beam detuning 
  %and the y-axis is the probe beam absorption and dispersion in the first row,
  %for Coupling beam in the second row and for Signal beam in the third row.These 
  %plots are for the case when all atoms are in level $|2\rangle$ as can be concluded by
   %looking at the rabi frequencies.}
%\label{fig:caseb22}
%\end{figure}
 \textbf {Case II:when probe and coupling have equal intensities or the CPT condition } 
For this case $b_1^*b_1 + b_2^*b_2$=1

We have put $\Delta\omega_{21}$=0 in the following expression.

%\textit{1.nonlinearity faced by the probe}
 %\begin{equation}
 %b_1^*b_3= \frac{2 \Omega_{13}^* \Delta \omega_{21} D^*}{D^*D+|\Omega_c|^2
 %|\Omega_{13}|^2}-\frac{\Omega_{13}^*|\Omega_{24}|^2(-|\Omega_{24}|^2 A^*
 %-|\Omega_c|^2
 %B^*)}{2(|-|\Omega_{24}|^2 A^* -|\Omega_c|^2 B^* |^2 +BB^*|\Omega_c|^2
 %|\Omega_{13}|^2)}
 %\end{equation}
 
 %\textit{2.nonlinearity faced by the coupling beam}
 
 %\begin{equation}
 %b_3b_2^* =\frac{-\Omega_c^* |\Omega_{24}|^2 |\Omega_{13}|^2 B^*}{2(
 %|-\Omega_{24}|^2 A^* -|\Omega_c|^2 B^* |^2
 %+BB^*|\Omega_c|^2
 %|\Omega_{13}|^2)}
 %\end{equation}
 
 \textit{nonlinearity faced by the Signal beam can be deduced from the following:}
 
 \begin{equation}
 b_4 b_2^*=\frac{\Omega_{24}^*|-|\Omega_{24}|^2 A^* -|\Omega_c|^2 B^* |^2B^*}{2(B^2B^{*2}|\Omega_c|^2 |\Omega_{13}|^2
 +BB^*|-|\Omega_{24}|^2 A^* -|\Omega_c|^2 B^* |^2)}
 \end{equation}

\begin{figure}
\includegraphics[scale=0.7]{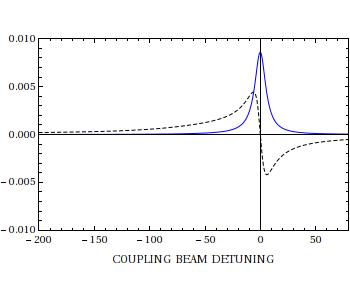} 
\caption{This figure shows the numerically calculated absorption and dispersion in the Signal beam for Case II.The x-axis and the line types used remain the same as in fig[\ref{fig:atom-laser1}]. The parameters are $\Omega_{13}$ = 0.1 MHz,  
  $\Omega_c$=0.1 MHz and $\Omega_{24}$=0.1 MHz.The analytic solution given in fig[\ref{fig:atom-laser6}]fits it the best. }
%The level scheme on the right is system2 and has been considered by Harris et. al.}
\label{fig:atom-laser5}
\end{figure}

\begin{figure}
\includegraphics[scale=0.7]{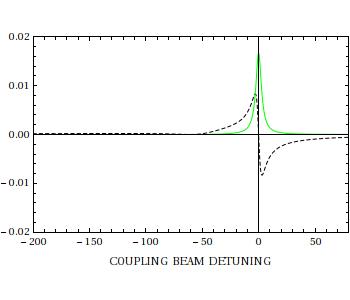} 
\caption{This figure shows the analytically calculated absorption and dispersion in the signal beam for Case II.The conditions remain the same as in fig[\ref{fig:atom-laser5}]. }
%The level scheme on the right is system2 and has been considered by Harris et. al.}
\label{fig:atom-laser6}
\end{figure}

We make a comparison with simulation when the rabi frequency of all the three fields is equal. We get the best agreement for the above solution, hence it is the valid one.The comparison with simulation is shown in fig.[\ref{fig:atom-laser5}] and [\ref{fig:atom-laser6}].
 
The nonlinearity here has an imaginary term that has $\Omega_{24}^*|\Omega_{24}|^4$ in the numerator. Squeezing of light will be hindered very little by the coexisting nonlinearity which has $\Omega_c$ in the numerator,because it is small and makes little difference.

\begin{figure}
\includegraphics[scale=0.7]{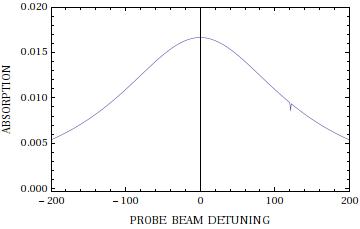} 
\caption{This figure shows numerically calculated absorption in the signal beam for Case II.The parameters remain the same as in fig[\ref{fig:atom-laser5}].The coupling is in resonance with the upper excited state,121 MHz away from the lower one.The atoms are in the uncoupled state so the linear absorption to an excited state is zero hence the lowest point of the CPT dip at probe detuning=121 MHz shows the strength of the nonlinearity that causes squeezing.The analytic solution shows the coexisting nonlinearity makes very little difference.Hence this nonlinear absorption that can cause squeezing is seen to be comparable to the linear absorption.}
%The level scheme on the right is system2 and has been considered by Harris et. al.}
\label{fig:atom-laser7}
\end{figure}

 Fig.[\ref{fig:atom-laser7}] shows  theoretical proof that this nonlinearity is comparable to linear features.

From the expressions above we see that different susceptibilies or combinations of
$b_a b_b^*$ describe different multiphoton processes belonging to type2
, when the susceptibilities are unequal. This can be possible because
the multiphoton process is visible in one of the beams of the system only, as it absorbs and
 emits a photon out of the beam , from the rest  of the beams it absorbs
and emits a photon, which replaces the absorbed photon exactly, which renders the process invisible. When the susceptibilities are equal, however, like in Case I,where the first terms in the expression for Coupling and Signal beams give
\begin{equation} 
|Im[\chi^{(5)}]|=|\frac{|\mu_{23}|^2 |\mu_{13}|^2 |\mu_{24}|^2 \Gamma_4/2}{2BB^*DD^*}|
\end{equation}
(We have used the Absolute value since the complex conjugate of $\chi^{(5)}$
gives the opposite sign,the two signs refer to absorption and emission), then we show in a later Section, the symmetry suggests  that the multiphoton process is type1 rather than type2.

%\textit{Note:How to decide whether the dipole matrix element $\mu_{ab}$ implies
%emission or absorption of a photon}

%Consider equation (\ref{sys1.3}).Consider the
%polarization due to an atom  $b_2^*\mu_{24}b_4$.
%We see whether Im[$b_2^*\mu_{24} b_4$] has a minus or plus sign.Since it has a plus
%sign $\mu_{24}$ implies absorption of a photon.The Imaginary part of complex conjugate 
%$b_2\mu_{24}^* b_4^*$ has a minus sign so $\mu_{24}^*$ implies emission of a
%photon.This should be found independently  for all $\mu$'s.

%       For an expression where the sign changes due to change in the sign of a
 %      detuning $\Delta\omega_{ab}$, as in system2, the interpretation of
 %      $\mu_{ab}$ will also change from denoting absorption to emission or vice
%       versa.

\section{The XPM nonlinearities in system2\label{sec:system2}}

Following Harris et. al.,we derive the following equations for
system2, from the Schrodinger equation, after making the rotating
wave approximation, in the steady state:

\begin{equation}
\Omega_{12} b_2 +\Omega_{13} b_3 =0
\end{equation}

\begin{equation}
\Omega_{12}^* b_1 - 2\Delta\tilde{\omega}_{21} + \Omega_{23} b_3 =0
\end{equation}

\begin{equation}
\Omega_{13}^* b_1 - 2\Delta\tilde{\omega}_{31} + \Omega_{23}^* b_2 =0
\end{equation}

where following Harris et. al. we define $\Omega_{12}$ =$\Omega_{1i}\Omega_{i2}$, where i is the intermediate
level, which is the rabi frequency corresponding to the fields $\omega_a$ and
$\omega_b$.$\Omega_{13}$ and $\Omega_{23}$ are the rabi frequencies corresponding to
the fields $\omega_d$ and $\omega_c$ respectively. To identify the fields look at
figure [\ref{fig:atom-lasersys2}].
Also, $\Delta\tilde{\omega}_{21}$=$\omega_{21}-\omega_a-\omega_b-\iota\Gamma_2/2$
and $\Delta\tilde{\omega}_{31}$=$\omega_{31}-\omega_d-\iota\Gamma_3/2$.
  %  It should be mentioned that the rotating wave approximation works for
    %this case only when $\omega_d$ and $\omega_c$ are taken to differ by
    %$\omega_a+\omega_b$.
    
    \textit{nonlinearity faced by the field $\omega_d$ can be deduced from the following:}

    \begin{eqnarray}
    b_1^* b_3 &=&
    \frac{\Omega_{23}^*\Omega_{12}^*(4\Delta\tilde{\omega}_{31}
   \Delta\tilde{\omega}_{21}-|\Omega_{23}|^2)^*}{|4\Delta\tilde{\omega}_{31}
   \Delta\tilde{\omega}_{21}-|\Omega_{23}|^2|^2} \nonumber \\ &+& \frac{2\Delta\tilde{\omega}_{21}\Omega_{13}^*}{4\Delta\tilde{\omega}_{31} \Delta\tilde{\omega}_{21}-|\Omega_{23}|^2}
\label{sys2.1}
    \end{eqnarray}
   
 The first term is the non linear part. The occurance of XPM fields in the form
 of single $\Omega$'s in the numerator $\Omega_{23}^*\Omega_{12}^*$ indicates
 the absorption or emission of a single photon each from all the fields
 $\omega_c$,$\omega_a$ and $\omega_b$ \cite{harris}.

    \textit{nonlinearity faced by the fields $\omega_a$ and $\omega_b$ can be deduced from the following:}
    
    \begin{eqnarray}                
    b_1^* b_2&=&
    \frac{\Omega_{13}^*\Omega_{32}(4\Delta\tilde{\omega}_{31}
   \Delta\tilde{\omega}_{21}-|\Omega_{23}|^2)^*}{|4\Delta\tilde{\omega}_{31}
    \Delta\tilde{\omega}_{21}-|\Omega_{23}|^2|^2} \nonumber \\&+&\frac{2\Delta\tilde{\omega}_{31}
    \Omega_{12}^*}{4\Delta\tilde{\omega}_{31} \Delta\tilde{\omega}_{21}-|\Omega_{23}|^2}
    \label{sys2.2}
    \end{eqnarray}
    The first term gives the nonlinear part.  
    
    \textit{nonlinearity faced by the field $\omega_c$ can be deduced from the following:}
    \begin{eqnarray}
    b_2^*b_3&=&-\frac{\Omega_{12}^*\Omega_{32}^*\Omega_{13}(4\Delta\tilde{\omega}_{31}
   \Delta\tilde{\omega}_{21}-|\Omega_{23}|^2)}{\Omega_{23}|4\Delta\tilde{\omega}_{31}
    \Delta\tilde{\omega}_{21}-|\Omega_{23}|^2|^2} \nonumber\\&+& 
   \frac{2\Delta\tilde{\omega}_{21}}{\Omega_{23}}|\frac{\Omega_{23}^*\Omega_{13}+
2\Delta\tilde{\omega}_{31}^*\Omega_{12}}{4\Delta\tilde{\omega}_{31}
    \Delta\tilde{\omega}_{21}-|\Omega_{23}|^2}|^2
   \label{sys2.3}
   \end{eqnarray}
   The first term gives the nonlinear part.
    To obtain the susceptibility ,$\chi^{(3)}$,$b_2^*b_3$ should be multiplied 
   with
   $\mu_{23}$ and not $\mu_{23}^*$. Only then the expressions in equations
   (\ref{sys2.1}),(\ref{sys2.2}) and (\ref{sys2.3}) represent the possible multi-photon
    processes

 $\chi^{(3)}(\omega_d,-\omega_a,-\omega_b,-\omega_c)$ or  
 $\chi^{(3)}(-\omega_d,\omega_a,\omega_b,\omega_c)$, 
 $\chi^{(3)}(\omega_c,\omega_a,\omega_b,-\omega_d)$ or 
 $\chi^{(3)}(-\omega_c,-\omega_a,-\omega_b,\omega_d)$, 
 $\chi^{(3)}(-\omega_a-\omega_b,\omega_d,-\omega_c)$ or
 $\chi^{(3)}(\omega_a+\omega_b,-\omega_d,\omega_c)$

 The complex conjugates $b_3^*b_1$,$b_2^*b_1$,$b_3^*b_2$ describe the reverse
multiphoton process.

 We get from the first terms of the three equations:
   \begin{equation}
   \chi^{(3)}=\frac{\mu_{23}^*\mu_{12}^*\mu_{13}(4\Delta\tilde{\omega}_{31}
   \Delta\tilde{\omega}_{21}-|\Omega_{23}|^2)}{|4\Delta\tilde{\omega}_{31}
    \Delta\tilde{\omega}_{21}-|\Omega_{23}|^2|^2}
    \end{equation}
     The absolute values of the first terms of Im[$\chi^{(3)}$]'s corresponding to the multiphoton
     processes $\chi^{(3)}(\omega_d,-\omega_a,-\omega_b,-\omega_c)$ or  
 $\chi^{(3)}(-\omega_d,\omega_a,\omega_b,\omega_c)$, 
 $\chi^{(3)}(\omega_c,\omega_a,\omega_b,-\omega_d)$ or 
 $\chi^{(3)}(-\omega_c,-\omega_a,-\omega_b,\omega_d)$, 
 $\chi^{(3)}(-\omega_a-\omega_b,\omega_d,-\omega_c)$ or
 $\chi^{(3)}(\omega_a+\omega_b,-\omega_d,\omega_c)$ 
  
are exactly
    equal.
% Though Harris et. al. have put $\chi^{(3)}(\omega_a ,\omega_b,-\omega_d,\omega_c)$ and  $\chi^{(3)}(\omega_b ,\omega_a,-\omega_d,\omega_c)$ separately to be equal,we think that might not be true.We give our reasons later in this section.We also discuss how this case of "partial" symmetry is useful.
The above nonlinearities, being equal, suggest type1 processes rather than
    type3. We show this in the next section. In type3 asymmetric processes the multiphoton process is visible in one beam only because there is a simultaneous reverse process that replaces the absorbed photons exactly with the emitted photons in the other beams. So the susceptibilities do not have to conform to any special symmetry condition, for the multiphoton process to belong to the type3. The same holds for type2. We note here that in type3 multiphoton processes it is not possible to have a reverse process that also has a emission out of the observed beam because, it is obvious, that changes the susceptibility of the system in a way that is not reflected in the calculations.(Then,the calculation shows only half the value of the actual susceptibility) So, the complex conjugate of susceptibility, for type3 processes, describes the same process instead of the reverse one.

  % The equality in Harris et.al.'s work\cite{harris} showing $\chi^{(3)}(\omega_a ,\omega_b,-\omega_d,\omega_c)$ = $\chi^{(3)}(\omega_b ,\omega_a,-\omega_d,\omega_c)$= $\chi^{(3)}(\omega_d,-\omega_a,-\omega_b,-\omega_c)$,might be only partially true because while  $\chi^{(3)}(\omega_a+\omega_b,-\omega_d,\omega_c)$ is symmetric with the other susceptibilities observed in $\omega_c$ and $\omega_d$, the multiphoton processes separately seen in the fields $\omega_a$ and $\omega_b$ can have an asymmetric part apart from the symmetric part that should necessarily be there.This case is important as it shows how an asymmetric susceptibility can be seen as a sum of symmetric and asymmetric part.
 
\section{Generation of cross field noise correlation due to the symmetry \label{sec:equalsus}}
 \textbf{ D=}  ${\tilde{\textbf{E}}}$ + $\varepsilon_0\textbf{P}$
           where $\tilde{\textbf{E}}$=E($\omega$)$e^{\iota\omega t}$

Symbols have the usual meaning.
            
        Energy density U in electic media=$\frac{1}{2}$(\textbf{D}.$\tilde{\textbf{E}^*}$)

     $\partial U/\partial t$ in steady state is a constant and is equal to the energy taken away from the field by absorptions by the atoms. When there is no medium this quantity is zero, as the energy outflow due to propagating waves,estimated by the Poynting vector, is compensated by the energy inflow originating in the laser current.

     $\omega_b$ in Schmidt et. al.'s system is the field under observation.

%\textbf{P}=N$\tilde{b}_4^*$$\mu_{24}^*$$\tilde{b}_2$$e^{-\iota\omega_{0b} t}$ where N is the atomic number density and  $\hbar\omega_{0b}$=$E_2$-$E_4$ where  $E_2$ and $E_4$ are energies of the levels marked as $|2\rangle$ and$|4\rangle$ in fig[\ref{fig:atom-lasersys1}].

%After making the rotating wave approximation,we have

%\textbf{P}=N(${b}_4^*$$\mu_{24}^*$${b}_2$$e^{\iota\omega_b t}$)$e^{-\iota\omega_{0b} t}$

   $\partial U/\partial t$=n$\hbar\omega_{b}$=Re[-$\frac{1}{2}\iota\omega_{b}$N${b}_4^*$$\mu_{24}^*$${b}_2$$E(\omega_b)$$e^{\iota\omega_b t}$] where n is the number of photons absorbed or emitted and $\partial E/\partial t$=0, in steady state.

Similarly, we find n for the other field $\omega_c$. We show that the multiphoton process described by the equal susceptibilities given in eq(\ref{eq:coupling}) and eq(\ref{sys1.3}) can be of type1 i.e., simultaneously visible in both the fields $\omega_c$ and $\omega_b$,and a generation of cross field noise correlation,\textit{For this we must show that the number of photons, n, absorbed or emitted from each field, is equal.}

  n$\hbar$=Re[-$\frac{1}{2}\iota$N${b}_2^*$$\mu_{23}$${b}_3$$E(\omega_c)$],for field $\omega_c$

(We actually mean only the first terms from the respective expressions of
${b}_2^*$$\mu_{23}$${b}_3$ and ${b}_4^*$$\mu_{24}^*$${b}_2$ since they only correspond to the multiphoton process we are considering)

We get
 \begin{equation}
n=|-\frac{N|\Omega_c|^2 |\Omega_{13}|^2 |\Omega_{24}|^2 \Gamma_4/2}{4 BB^*DD^*}|
\label{equality} 
\end{equation}
  for both the fields, $\omega_c$ and $\omega_b$
 This can similarly be proved for all cases of the symmetry.
%We note here that the multiphoton processes seen here can be type1,because the field energy has the form ($\chi^{(x+y+z)} E_1^{x}E_2^{y}E_3^{z}$)$E_1$ where x+1=y=z and $E_1$ is the observed field.This is a sufficient but not necessary condition for a type1 multiphoton process.

      %     The necessary condition is obvious from the above derivation.Say ($\chi^{(x+y+z)} E_1^{x}E_2^{y}E_3^{z}$)$E_1$ is the form of field energy for field$E_1$ and say  ($\chi^{(r+s+t)} E_1^{r}E_2^{s}E_3^{t}$)$E_2$ is the form of field energy for field $E_2$.Then x+1=r,y=s+1,z=t is the necessary condition.
%Actually,the susceptibilities occuring in this work have such a form that susceptibilities being equal automatically ensures the above condition is met.

\section{XPM Nonlinearity for the case of system3\label{sec:system3}}
The consideration of this system leads to  a postulation of existence of type3 multiphoton processes.The fig.[\ref{fig:atom-lasersys3}] shows the relevent level scheme. After making the rotating wave approximation,we get the following equations:
  \begin{equation}
  -2 \Delta\omega_{21} \Delta b_2 +\Omega_{23}\Delta b_3 +\Omega_{24}\Delta b_4+
  \Omega_{w23} b_3 + \Omega_{w24}b_4=0
  \end{equation}    
     
 \begin{equation}
 \Omega_{31}^*\Delta b_1 +\Omega_{32}^*\Delta b_2 -2A\Delta b_3 +\Omega_{w23}^*b_2=0
 \end{equation}
 
\begin{equation}
 \Omega_{42}^*\Delta b_2 -2B\Delta b_4 +\Omega_{w24}^*b_2=0 
 \end{equation}
 
 Since we use the results  from system1 in solving for the nonlinearity, the rotating wave approximation should
 be the same as before.$\Delta b$'s are the perturbation probability amplitudes due to the weak couplings.

we solve the equations for the form $ b_a^*\Delta b_b$ using the conditions
   \begin{equation}
   (b_1 + \Delta b_1)(b_1^* + \Delta b_1^*)+(b_2 +\Delta b_2)(b_2^* +\Delta
   b_2^*)=1
   \end{equation}
  where $b_1 b_1^*$=1 and 
  %we also use the following condition
  %\begin{equation}
 %\Delta b_1 \Delta b_1^* + \Delta b_2 \Delta b_2^*
  %\end{equation}
% which means that  
  
\begin{equation}
 b_1^* \Delta b_1 +b_2^* \Delta b_2=0 
 \end{equation}
   
 Solving the equations ,at two photon detuning
 zero, for $ b_2^*\Delta b_4$,one of the solutions we get is :
 \begin{eqnarray}
 [2|\Omega_{31}|^2|\Omega_{23}|^2B 
 +2BDD^*]b_2^*\Delta b_4 \nonumber \\ &=&
 +\frac{\Omega_{42}^*\Omega_{w23}^*\Omega_c|\Omega_c|^2|\Omega_{13}|^2}{D}    
  +\frac{\Omega_{w24}^*|\Omega_c|^4|\Omega_{13}|^4}{|D|^2} \nonumber \\ &+&
  \frac{2\Omega_{42}^*\Omega_{w24}\Omega_{24}^*|\Omega_c|^2|\Omega_{13}|^2A}{BD} 
  -\frac{\Omega_{24}^*\Omega_c^*\Omega_{w23}|\Omega_{24}|^2|\Omega_{13}|^2A}{BD}
  \nonumber \\ &-& 
  \frac{\Omega_{w24}^*|\Omega_c|^4\Omega_{13}^2}{D}  
  \end{eqnarray}
  
For $b_4^* \Delta b_2$, one of the solutions we get is:
  \begin{eqnarray}
 [2|\Omega_{31}|^2|\Omega_{23}|^2BB^* 
 +2BB^*DD^*]b_4^*\Delta b_2 \nonumber \\ &=&
 \frac{\Omega_{42}\Omega_{w24}^*\Omega_{24}A|\Omega_c|^2|\Omega_{13}|^2}{D}    
  +\frac{\Omega_{24}\Omega_{w23}^*\Omega_cB|\Omega_c|^2|\Omega_{13}|^2}{D}
  \nonumber \\ &+&
  \frac{2\Omega_{42}\Omega_{w23}\Omega_c^*AB|\Omega_c|^2|\Omega_{13}|^2}{BD} 
  \nonumber \\ &+&
  \frac{\Omega_{24}^*\Omega_{24}^*\Omega_{w24}AB|\Omega_c|^2|\Omega_{13}|^2}{BD}  
  \end{eqnarray}
  All solutions including the ones seen in $\Omega_{w23}$ field are similar to the extent that they show a ninth order nonlinearity .
%and type3 multiphoton processes.We see from the simulation\cite{andal} done for system3 that an identical solution is not found in the $\Omega_{w23}$ field.Hence it is confirmed that the system3 shows multiphoton processes of type3 rather than type1.
   We note the occurance of two $\Omega_{24}$'s or $\Omega_{24}^*$'s in the expression
   without its conjugate counterparts which would look like two consecutive photons getting
   absorbed(emitted) from the field. But it is accompanied with an emission(absorption) in the weak field,so it is not absorbing the second photon and going to a higher virtual level.

       Since the nonlinearity absorbs photons from four different fields it is particularly sensitive to cross field correlations.Though the nonlinearities in system2 absorb photons from only three fields, they
have the qualitative difference that they in no cases, at the same time require the absorption of more than one photon from a single field. 
These ninth order susceptibilities are shown by the analytic calculation to be completely asymmetric and belong to type2 because of reasons given earlier.Also,we remind the reader,we are making calculations for zero velocity atoms.
\section{Conclusions \label{sec:end}}
To summarize,we postulate or conjecture three types of multiphoton process. While the type of multiphoton processes seen here are shown,to be only type2 
% or type3(that are visible only in one field)
when the susceptibility terms are asymmetric, when the susceptibilities faced by two fields show symmetry, then the multiphoton process is type1(that is visible in two or more fields simultaneously) rather than type2.
  Type3 is shown to not exist. Through these arguments we have seen a way to pick up the valid solution from the many solutions that Harris et. al.'s formalism often gives in system2.
%Using our analysis here ,we only partially agree with the first example of the symmetry, that Harris et.al.\cite{harris} give without going into the origin of this symmetry but find that the example is useful in deciding how an asymmetric case can be a sum of symmetric and asymmetric parts.
We have shown that the symmetry generates cross field noise correlation.Type1 processes have the possibility of leading to strong sources of temporally entangled photons, of different frequency, as the nonlinearities are EIT enhanced. We have also shown, a $\chi^{(5)}$ XPM nonlinearity that can be a promising source of light squeezing as it is comparable to the linear absorption. It squeezes due to the Imaginary part of the susceptibility.We show this is unusual so this source is qualitatively novel.In the process of showing the suggestion of existence of type2 multiphoton processes, using a modified system, we show a way to generate an easily observable $\chi^{(9)}$ absorption resonance, which being  a high order nonlinearity is expected to be very sensitive to cross field correlations.We have done an experiment described in our previous work\cite{andal} but the order of nonlinearity newly interpreted, that shows the $\chi^{(9)}$ feature,only an order smaller than linear absorption.
%We have shown how this nonlinearity can be used to test the conjucture of type2 and type3 processes.
\section{Acknowledgements}
I would like to thank  Prof. Hema Ramachandran  and Prof.Andal Narayanan mainly for the background work, to this paper, done prior to year 2008 . I also thank Andal Narayanan for help with simulations and the experimental plot shown in this work. I thank Prof.
R. Srikanth for many insightful discussions. I thank Prof. Reji Philip
and Manukumar for discussions. I thank Prof. Astrid Lambrecht for guiding me
to important references.

\end{document}